\documentclass[showpacs,aps,preprint,graphicx,psfig]{revtex4}
\usepackage{amsmath}
\usepackage{graphicx}
\usepackage{amssymb}
\usepackage{bm}
\usepackage{color}

\begin{document}
\title{Main restrictions in the synthesis of new superheavy elements: quasifission or/and fusion-fission}
\author{A. K. Nasirov$^{1,2}$, Kyungil Kim$^3$, G. Mandaglio$^{4,5,6}$,   G. Giardina$^{4,5}$,
A. I. Muminov$^2$,  Youngman Kim$^3$}

\affiliation{
$^1$Joint Institute for Nuclear Research, Joliot-Curie 6, 141980 Dubna, Russia\\
$^2$Institute of Nuclear Physics, Ulugbek, 100214, Tashkent, Uzbekistan\\
$^3$Rare Isotope Science Project, Institute for Basic Science, Daejeon 305-811, Republic of Korea\\
$^4$Dipartimento di Fisica e di Scienze della Terra  dell' Universit\`a di Messina,
Salita Sperone 31, 98166 Messina, Italy\\
$^5$Istituto Nazionale di Fisica Nucleare, Sezione di Catania,  Italy\\
$^6$Centro Siciliano di Fisica Nucleare e Struttura della Materia 95125 Catania, Italy \\
}
\begin{abstract}
The synthesis of superheavy elements stimulates the effort to
study the peculiarities of the complete fusion with massive
nuclei and to improve theoretical models in order to extract
knowledge about reaction mechanism in heavy ion collisions at low
energies. We  compare the theoretical results of
the compound nucleus (CN) formation and evaporation residue (ER) cross sections obtained for
the $^{48}$Ca+$^{248}$Cm and $^{58}$Fe+$^{232}$Th reactions leading to the formation
of the isotopes $A=296$ and $A=290$ of the new superheavy element Lv ($Z=116$),
 respectively.
The ER cross sections, which can be measured directly, are
determined by the complete fusion and survival probabilities of the
heated and rotating CN. Those probabilities can not be
 measured unambiguously but the knowledge about them is important to
study the formation mechanism of the observed products and to estimate
the ER cross sections of the expected isotopes of elements.
For this aim, the $^{48}$Ca+$^{249}$Cf and $^{64}$Ni+$^{232}$Th reactions
are considered too.
  The use of the mass values of superheavy nuclei calculated in the
 framework of the macroscopic-microscopic model by Warsaw
 group  leads to smaller ER cross section
 for all of the reactions
in comparison with the case of using the masses
 calculated by  Peter  M\"oller  {\it et al}.
\end{abstract}
\pacs{25.70.Jj, 25.70.Gh, 25.85.-w}
\date{Today}
\maketitle
\section{Introduction}
\label{intro}
The heaviest superheavy element which was synthesized in the cold-fusion
($^{70}$Zn+$^{209}$Bi) reaction was $Z$=113: it was observed with a cross
section of tens femtobarn, {\it i.e.} 55 fb \cite{RIKEN}.
The hot fusion reactions were more favorable to synthesis of
superheavy elements $Z=114-118$ in Joint Institute for Nuclear Research
(Dubna, Russia) \cite{Oganessian04,FLNR}. As a result, the Mendeleev Table of chemical elements
has been extended by new elements Fl ($Z=114$) and Lv ($Z=116$) after those experimental
results have been certificated in the other experiments devoted to synthesis
of elements 114 and 116 in the $^{48}$Ca+$^{244}$Pu \cite{LBNL114}
and $^{48}$Ca+$^{248}$Cm \cite{GSI116,SHofmann116} reactions, respectively.

The very small cross sections of synthesis of superheavy elements
require to find favorable reactions (projectile and target pair) and the
optimal beam  energy range which is very narrow (no more 10 MeV).
It is necessary to establish conditions to increase  the events
of the ER formation.
The experimental way to estimate possibility of the synthesis of the
wanted superheavy element is to study yield of fissionlike products
and their mass-angle distributions. The presence of the mass-symmetric or
around mass-symmetric binary reaction products and their isotropic angular
distribution is a necessary condition. The CN formation
must survive against fission to be registered as the evaporation residue.
The fission probability can be very large due to decrease or disappearance
of the  fission barrier at large values of the excitation  and
rotational energies of the just formed CN. So, survival probability of
the heated and rotating CN against fission is also an important factor
which is subject for the further theoretical and experimental studies.
Therefore, theoretical studies of the  reaction mechanism in collision
of massive nuclei and of the range of the favorable beam energy are essential.
\begin{figure*}
\begin{center}
\resizebox{0.75\textwidth}{!}{\includegraphics{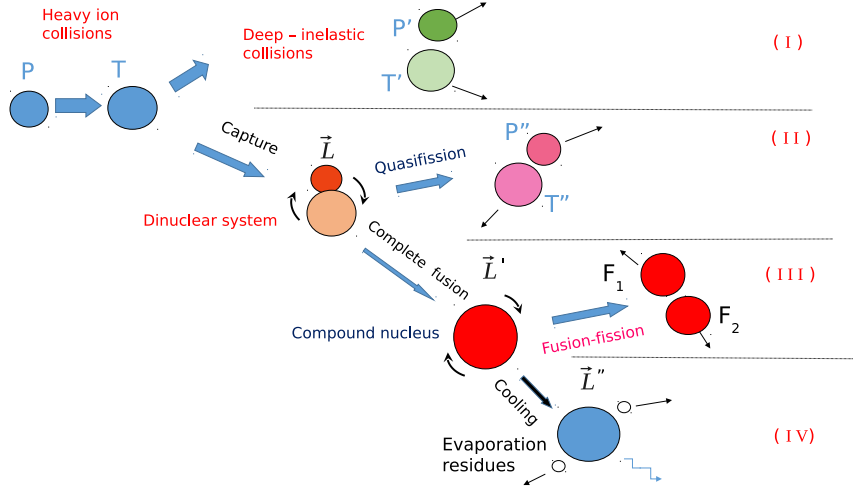}}
\caption{\label{stages} (Color online) The sketch of the damped reaction
channels (I-III) leading to formation of binary or fissionlike fragments which compete in the
way to complete fusion (IV) of the initial projectile P and target T nuclei:
P' and T' are deep-inelastic collision (not full momentum transfer) products;
P'' and T'' quasifission (full momentum transfer) products; F$_1$ and F$_2$ are fusion-fission
fragments. }
\end{center}
\end{figure*}

\begin{figure}
\begin{center}
\resizebox{0.57\textwidth}{!}{\includegraphics{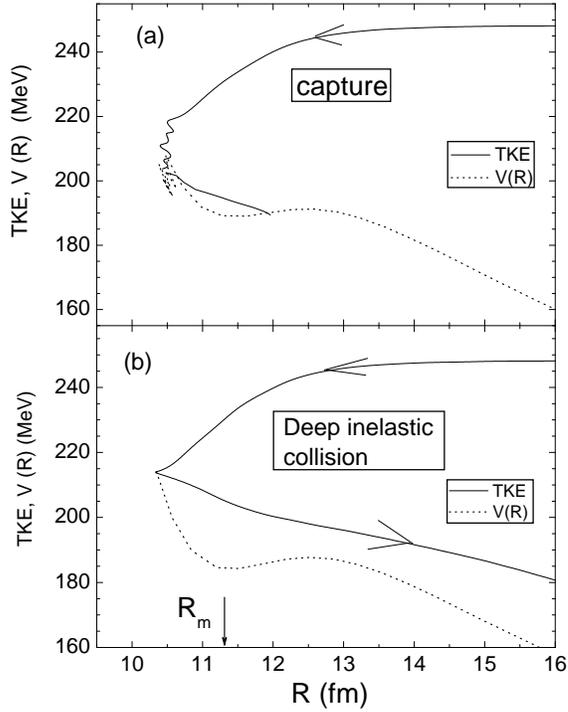}}
\vspace*{-2.55 cm} \caption{\label{TrapWell}
Illustration of capture (a) and deep inelastic collision (b) at heavy ion collisions.
Total kinetic energy (TKE) of the relative motion and the part of nucleus-nucleus potential
are shown by solid and dotted curves, respectively.}
\end{center}
\end{figure}

The ER formation process is the last stage of the reaction mechanism
in heavy ion collisions near the Coulomb barrier energies.
The sketch of the preceded stages to formation of the evaporation
 residues is presented in Fig. \ref{stages}.
Naturally, the threshold interaction of projectile and target nuclei ending with
the ER formation begins with the multinucleon transfer reactions.
We should distinguish two kinds of reactions which are in competition
during heavy ion collisions at low (near the Coulomb barrier) energies
of the projectile.
The first one is deep-inelastic collisions and the second one is
capture reactions shown on the top of Fig. \ref{stages}.
The difference between capture and deep-inelastic collision is
 whether a path of the relative motion has been trapped into the well of the
 nucleus-nucleus interaction or not (see Fig. \ref{TrapWell}).
In the case of capture the full momentum transfer takes place, while it does not
 in the deep-inelastic collisions
(I-channel of binary reaction in heavy ion collisions, see Fig. \ref{stages}).
In both cases the dinuclear system (DNS) will be formed.
 The lifetime of the DNS formed in the capture process will be sufficiently long.
  Two conditions must be satisfied for capture:
1) the initial energy  $E_{\rm c.m.}$  of projectile in the center-of-mass system
should be sufficiently large to overcome the interaction barrier (Coulomb
barrier + rotational energy of the entrance channel),
2) some part of the relative kinetic energy has to be dissipated in order that
  DNS would be trapped in the well of the
nucleus-nucleus interaction potential \cite{GiaEur2000}. If there is not a potential
well the deep-inelastic collision takes place only.
Here we state that mass distributions of the deep-inelastic collision and capture reaction
 (quasifission) products may overlap in reactions with nuclei having magic proton
or/and neutron numbers \cite{FazioMPL2005}. In this case, usually, the contribution
of quasifission is arrogated to the  deep-inelastic collision.

The capture is the first step in the way to ER formation, and it leads
to formation of a  molecule-like DNS which can evolve by changing its charge and mass asymmetry
due to multinucleon transfer and by changing its shape. The angular  velocity of the
 rotating  DNS depends on the initial value of orbital angular momentum which is
determined by the impact parameter $b$ and momentum $P$ of the collision:
$\vec{L}=\vec{\ell}\hbar=[\vec{b} \times \vec{P}]$.

The study of dynamics of the heavy ion collisions near the Coulomb barrier
energies showed that complete fusion does not occur immediately
in the case of  massive nucleus collisions
\cite{Back32,VolPLB1995,AdamianPRC2003,FazioEPJ2004,NasirovNPA759}.
In Ref. \cite{Wang85}, authors estimated the excitation functions
of the ER formation in the $^{50}$Ti+$^{249}$Bk and $^{50}$Ti+$^{249,252}$Cf reactions
by using a newly developed DNS model with a dynamical potential energy surface.
One of interesting conclusions of authors is that taking into account dynamical deformation
of DNS fragments leads to the decrease of the fusion probability and, consequently,
to the decrease of the ER formation cross section. But authors
of Ref. \cite{Wang85} restricted by analysis of the collision  with the tip-tip orientation
of the two deformed nuclei.

Evolution of DNS may end with the CN formation in the equilibrium state
or may arrive at two fragments without reaching the CN state.
Quasifission takes place when DNS prefers to break down into two fragments
instead of being transformed into fully equilibrated  CN.

For example, the total kinetic energy of the quasifission fragments is close to that of fission fragments.
The mass and angular distributions of the fragments depend on the entrance channel properties
and may overlap, causing difficulties in identification of a mechanism which produces
the corresponding reaction products \cite{NasirovPRC79}. We should stress the dominant
role of the quasifission channel in reactions
with massive nuclei that causes strong hindrance to the formation of CN during the
evolution of DNS. This important topic will be discussed in the present paper.

Mononucleus survived  against quasifission can reach equilibrium shape, and it may be
transformed into CN. The heated and rotating CN must survive against
 fission  to be registered in a detector as the recoil nucleus through
de-excitation by emission of neutrons, protons, $\alpha$-particles, and
gamma-quanta.
The fusion-fission forms binary fragments (the channel (III) in Fig. \ref{stages}).
 This fusion-fission channel can be enriched by
the fast fission products. Fast fission, which requires no fission barrier,
 is splitting of mononucleus before reaching
equilibrated CN. The properties of fission barrier can be described by liquid drop model
for the intermediate mass and heavy nuclei.
 The stability  of very heavy atomic nuclei with charge number $Z >$ 106
 against fission appears only due to shell effects in their binding energy
\cite{ArmbrusAnnRevNP,Sobiczewski2007}, which  is sensitive to the
 $\ell$ and $E^*_{\rm CN}$ values.
  Fission barrier of nucleus related with both nature (liquid drop and shell effect)
 disappears at large values of angular momentum
$L$ \cite{Sierk} and when shell effects  have been completely damped \cite{MandaglioPRC86}
as a function of the excitation energy and angular momentum.
 The increase of the  angular momentum leads
to damping of the shell effects and a mononucleus which has survived against quasifission
can not reach CN equilibrium shape and suffers  fast fission.
At last, the CN, which survived against fission by emission of  particles
 during its cooling,  forms the ER which is a product of channel (IV) in Fig. \ref{stages}.

  The main  scope of this work is to reproduce the measured data
 for the superheavy elements with $Z=116$ and $Z=118$ formed in the
 $^{48}$Ca+$^{248}$Cm and $^{48}$Ca+$^{249}$Cf reactions, respectively, and
to make predictions for  $\sigma_{\rm ER}$ in the
 $^{58}$Fe+$^{232}$Th and $^{64}$Ni+$^{232}$Th reactions which can
 be used in future experiments.

\section{Outline of the approach}
\label{sec:1}

\begin{figure}
\begin{center}
\resizebox{0.50\textwidth}{!}{\includegraphics{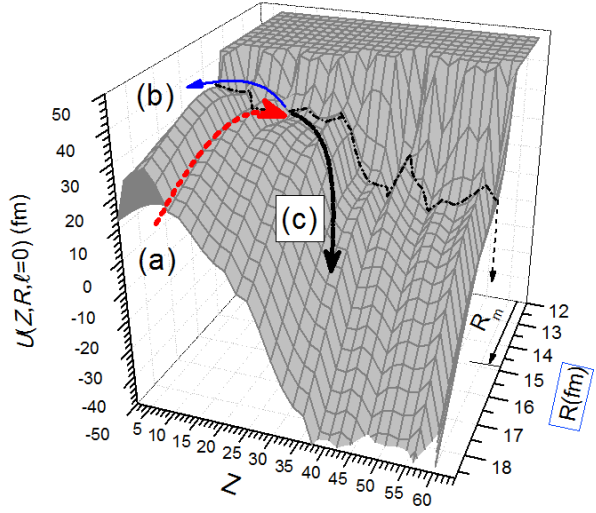}}
\vspace*{-0.10 cm} \caption{\label{PESCaCm} (Color online) Potential energy surface
 calculated for the DNS leading to formation of the  $^{296}$116 CN
 as a function of the relative distance between the centers of mass
of interacting nuclei and mass number of a fragment. The capture stage path is shown
by arrow (a) and complete fusion by multinucleon transfer occurs (b) if system overcomes
intrinsic fusion barrier. Arrow (c) shows one of possibilities of the DNS quasifission.
The broken dot-dashed line corresponds to the driving potential $U(Z,R_m)$ which is determined
by the minimum values of the potential wells for each charge value $Z$. $R_m$ is the position
of the minimum value of interaction potential on the relative distance $R$.
 }
\end{center}
\end{figure}

\begin{figure}
\hspace*{-0.5cm}\resizebox{0.54\textwidth}{!}{\includegraphics{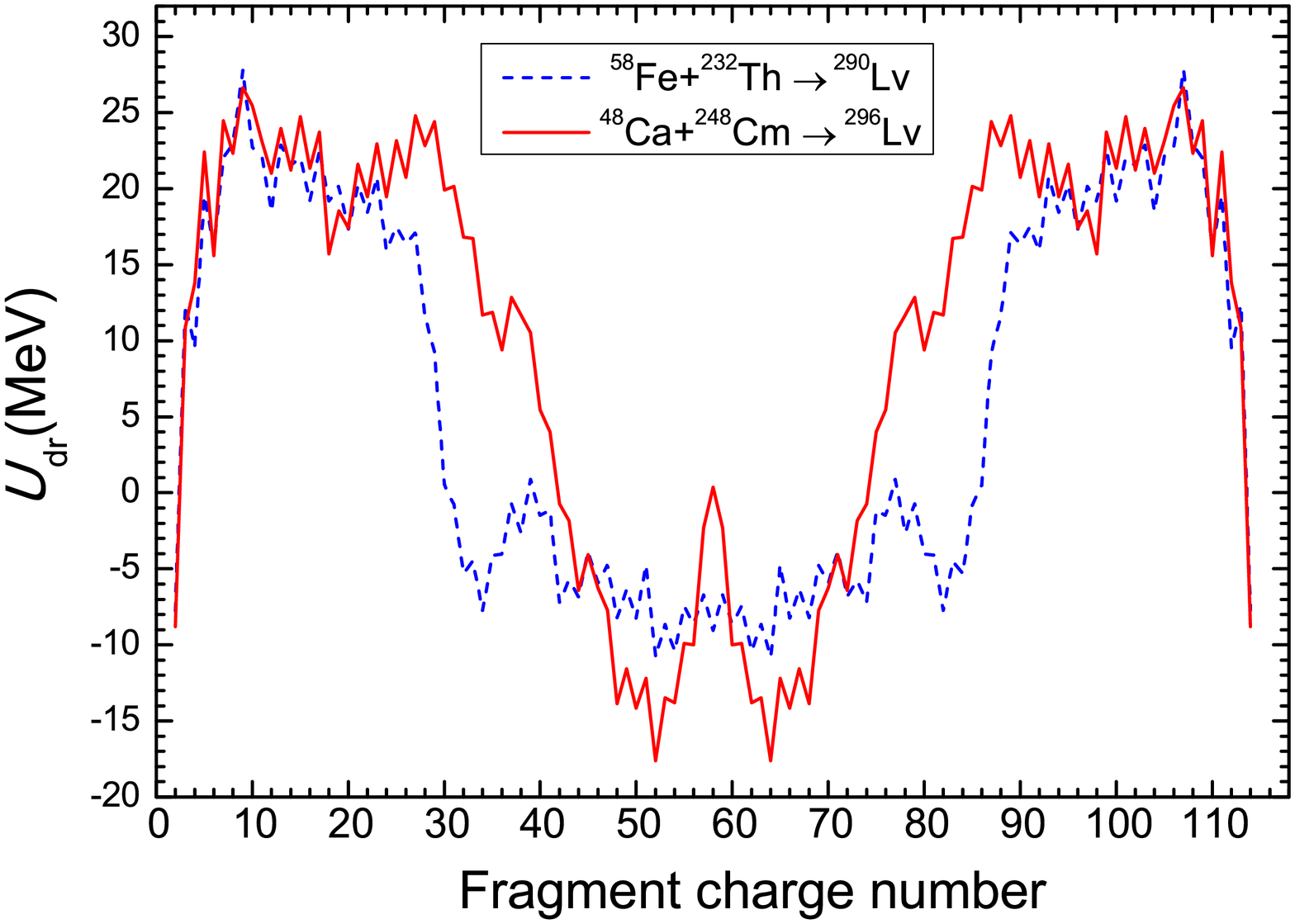}}
\vspace*{-0.80 cm} \caption{\label{DrivComp1} (Color online) Comparison of the driving
potentials calculated for the DNS formed in the $^{48}$Ca+$^{248}$Cm and $^{58}$Fe+$^{232}$Th
reactions which can lead to formation of isotopes $A$=296 and 290 of new superheavy element
Lv ($Z$=116) as a function of the fragment charge number.}
\end{figure}

 The dynamics of heavy ion collisions at low energies is determined by the
 peculiarities of the nucleus-nucleus interaction.
 The landscape of  potential energy surface (PES) $U$ plays a main role in an estimation
  of the complete fusion probability in competition with quasifission. It is calculated
 as a sum of the reaction energy balance ($Q_{\rm gg}$) and the nucleus-nucleus
 potential  ($V(R)$) between interacting nuclei:
\begin{equation}
\label{PESeq}
U(Z,A,\ell,R)=Q_{\rm gg}+V(Z,A,R,\ell),
\end{equation}
where $Z=Z_1$ and $A=A_1$ are charge and mass numbers of a DNS fragment
while the ones of another fragment are $Z_2=Z_{\rm tot}-Z_1$ and $A_2=A_{\rm tot}-A_1$,
where $Z_{\rm tot}$  and $A_{\rm tot}$ are the total charge and mass numbers of a reaction,
 respectively; $Q_{\rm gg}$ is the reaction energy balance
used to determine the excitation energy of CN:
$Q_{\rm gg}=B_1+B_2-B_{\rm CN}$. The binding energies
the initial projectile and target nuclei ($B_1$ and $B_2$)
are obtained from the mass tables in Ref. \cite{MassAW95}, while the one
of CN  ($B_{\rm CN}$) are obtained from the mass tables
\cite{MolNix,Muntian03}
The use of nuclear binding energies including shell effects in
the PES and driving potential of DNS leads to the appearance of hollows
on the PES around the charge and mass symmetries corresponding to the constituents of
DNS with the magic proton or/and neutron numbers
 (see Figs. \ref{PESCaCm} and \ref{DrivComp1}).

The driving potential $U_{\rm dr}(Z,R_m)$ is shown by the broken dot-dashed line in
Fig. \ref{PESCaCm}
and it is  determined by the minimum values of the potential wells for each charge value $Z$.
The position of the minimum value of interaction potential on the relative distance
 is denoted as $R_m$. The values of $U_{\rm dr}(Z,R_m)$ as a function of angular momentum
$\ell$ are found from the data of PES calculated by formula
\begin{equation}
U_{\rm dr}(Z,A,\ell,R_m)=Q_{\rm gg}+V(Z,A,R_m,\ell).
\end{equation}
 If there is no potential well of $V(Z,A,R,\ell)$ at large values of angular momentum
 or for symmetric massive nuclei, we use  $R_m$ corresponding to the smallest value of
the derivation $\partial V(Z,A,R_m,\ell)/\partial R$ in the contact area of nuclei.

 Capture stage is shown by arrow (a) in Fig. \ref{PESCaCm} and  one of possibilities
 of the DNS quasifission is shown by (c), while arrow (b) shows complete fusion direction.

In  Fig. \ref{DrivComp1} we compare the driving
potentials calculated for the DNS formed in the $^{48}$Ca+$^{248}$Cm and $^{58}$Fe+$^{232}$Th
reactions which can lead to the CN formation  with $^{296}$Lv and $^{290}$Lv,
respectively,  as a function of the fragment charge number. PES and driving potential are function
of the orientation angles $\alpha_1$ and $\alpha_2$ of interacting nuclei. The figures
\ref{PESCaCm} and \ref{DrivComp1} have been calculated for the intermediate values
of the orientation angles.

It is seen from Fig. \ref{DrivComp1} that the driving potential decreases
abruptly for the fragment with charge number larger than $Z=$30 for
both the $^{48}$Ca+$^{248}$Cm and $^{58}$Fe + $^{232}$Th reactions.
If the value of the driving potential corresponding
 to the entrance channel is very low relative to its maximum value in the
fusion direction $Z\rightarrow$0, the intrinsic fusion barrier $B^*_{\rm fus}$
becomes larger and the hindrance to complete fusion will be very strong.
It is determined as a difference between the maximum value
of the driving potential between $Z=0$ and $Z=Z_P$ and its value
corresponding to the initial charge value:
\begin{equation}
 B^*_{\rm fus}=U_{\rm driv}^{\rm max}- U_{\rm driv}(Z_P),
\end{equation}
where $U_{\rm driv}^{\rm max}=U_{\rm driv}(Z=9)$;  $Z_P=20$ and $Z_P=26$
for these two reactions under discussion.
The values of  $B^*_{\rm fus}$ are
equal to 8.5 MeV and 11.5 MeV for the
 the $^{48}$Ca+$^{248}$Cm and $^{58}$Fe+$^{232}$Th reactions, respectively
(see Fig. \ref{DrivComp1}).

 The mass and charge asymmetry of the entrance channel is enough large
  in the reactions with $^{48}$Ca-projectile and actinide-targets,
 and so  we deal with hot fusion reactions: a CN is formed with relatively
large excitation energy and with large fusion probability.
 To form DNS with the target nucleus,
 a projectile must overcome the entrance channel barrier consisting of the
 Coulomb potential and rotational energy moving along the relative distance $R$.
 As we can see from Fig. \ref{PESCaCm} the maximum value of PES along  the axis $R$
 for the fixed value $Z=20$ is larger than the one for the charge asymmetry
 $Z=26$. Therefore,  the large excitation energy of CN
 is an inevitable circumstance in the hot fusion
 reactions because after capture and formation of the DNS, the value of PES
 corresponding to the charge asymmetry of entrance channel is settled at  higher
 points of its hollow in comparison with the case of cold fusion reactions.
 Therefore, even  if the CN is formed by as possible minimum
beam energy, it is excited at energies  higher than $E^*_{\rm CN}$=30 MeV.

The values of the potential energy surface calculated for the
$^{48}{\rm Ca}$+$^{248}{\rm Cm}$ reaction corresponding to the
initial charge numbers $Z_P$=20 and  $Z_T$=96 are placed in the
valley formed due to shell effects (see solid line in Fig. \ref{DrivComp1}).
Note that $^{48}{\rm Ca}$ is double magic nucleus ($Z=20$ and $N=28$).
 As a result there is an intrinsic fusion barrier $B^*_{\rm fus}$
which must be overcome by DNS to be transformed into CN.

The capture stage path is shown
by arrow (a) and complete fusion by multinucleon transfer occurs (b) if system overcomes
intrinsic fusion barrier. Arrow (c) shows one of possibilities of the DNS quasifission.
The broken dot-dashed line corresponds to the driving potential $U(Z,R_m)$ which is determined
by the minimum values of the potential wells for each charge value $Z$. $R_m$ is the position
of the minimum value of interaction potential on the relative distance $R$.

The advantage of hot fusion reactions in comparison with cold fusion
reactions is connected with the relatively small hindrance in the stage of
CN formation.

\begin{figure}
\hspace*{-0.5cm}
\resizebox{0.535\textwidth}{!}{\includegraphics{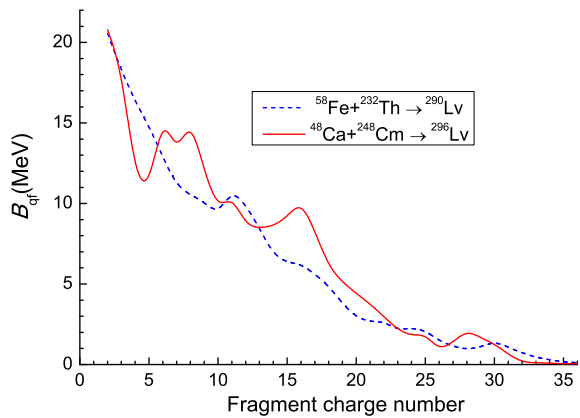}}
\vspace*{-0.35 cm} \caption{\label{Bqfis} The quasifission barriers of the
DNS fragments as a function of their charge numbers for the $^{48}{\rm Ca}$+$^{248}{\rm Cm}$
(solid curve) and  $^{58}{\rm Fe}$+$^{232}{\rm Th}$ (dashed curve) reactions.}
\end{figure}

In the case of the $^{58}{\rm Fe}$+$^{232}{\rm Th}$ reaction, the value of driving potential
corresponding to the initial charge numbers $Z_P$=26 and $Z_T$=90 is farther and lower
from the fusion barrier placed at $Z=9$ (see dashed line in Fig. \ref{DrivComp1}).
 It is much lower in comparison with that for
the $^{48}{\rm Ca}$+$^{248}{\rm Cm}$ reaction.
 This means that the intrinsic fusion
  barrier for the $^{58}{\rm Fe}$+$^{232}{\rm Th}$ reaction is larger. Furthermore,
  the quasifission barrier providing relative stability of DNS formed in this
  reaction is smaller in comparison with that obtained for
  the $^{48}{\rm Ca}$+$^{248}{\rm Cm}$ reaction.
  This comparison is presented in Fig. \ref{Bqfis}. As a result the
  fusion probability for the $^{58}{\rm Fe}$+$^{232}{\rm Th}$ reaction should be small.

We calculate the ER cross section at the given energy
as a sum of the contribution of reaction
channels corresponding to the different partial waves:
\begin{equation}
\sigma_{\rm ER}^{(x)}(E^{(x)}_{\rm CN})=\Sigma^{\ell_d}_{\ell=0}
(2\ell+1)\sigma_{\rm ER}^{(x)}(E^{(x)}_{\rm CN},\ell),
\end{equation}
where $\sigma_{ER}^{(x)}(E^{(x)}_{\rm CN},\ell)$ is the partial cross section of ER
formation with excitation energy $E^{(x)}_{\rm CN}$ after each step $x$
of the de-excitation cascade after the emission from the hot CN
of particles $\nu(x)$n + $y(x)$p + $k(x)\alpha$ + $s(x)$
(where $\nu(x)$, $y$, $k$, and $s$ are numbers of neutrons, protons,
$\alpha$-particles, and $\gamma$-quanta) by formulae
(See Refs. \cite{GiaEur2000,NasirovNPA759}):
\begin{equation}
\sigma_{\rm ER}^{(x)}(E^{(x)}_{\rm CN},\ell)=
\sigma^{(x-1)}_{\rm ER}(E^{(x)}_{\rm CN},\ell)W^{(x-1)}_{\rm sur}(E^{(x-1)}_{\rm CN},\ell),
\end{equation}
where  $\sigma^{(x-1)}_{\rm ER}(E^{(x-1)}_{\rm CN},\ell)$ is the partial formation cross section
of the excited intermediate nucleus of the $(x-1)$th step and
$W^{(x-1)}_{\rm sur}$ is the survival probability of the
$(x-1)$th intermediate nucleus against fission along the
de-excitation cascade of CN; obviously
$$\sigma^{(0)}_{\rm ER}(E^*_{\rm CN},\ell)=\sigma_{\rm fus}(E^*_{\rm CN},\ell)$$
{\it i.e.} the first evaporation  starts from the heated and rotating CN
and $E^{(0)}_{\rm CN}=E^*_{\rm CN}=E_{\rm c.m.}+Q_{\rm gg}-V_{\rm rot}(\ell)$;
$V_{\rm rot}(\ell)$ is rotational energy of CN.

Due to the dependence of the fission barrier on the angular momentum $\ell$,
the survival probability  $W_{\rm sur}(E^*_{\rm CN},\ell)$ depends on $\ell$.
The damping of the shell corrections determining the fission barrier is
taken into account as in Ref. \cite{MandaglioPRC86}.

If the colliding nuclei are deformed, the possibility of collision with
arbitrary orientation angles of their symmetry axis
 should be considered.
Due to the dependencies of the nucleus-nucleus potential ($V$) and moment of
inertia for DNS ($J_R$) on the orientation angles  of  the  axial
symmetry axis of the  deformed  nuclei,  the excitation function  of
 the  capture  and  fusion  are  sensitive  to  the values of  orientation angles.

In the case of spherical nuclei we can take into account of the vibrational
excitation of their surfaces due to interactions. Here this procedure
is used  for the projectiles $^{48}$Ca and $^{64}$Ni.

\subsection{Capture and fusion cross section in collisions of deformed nuclei}

The final results of the partial capture and
 complete  fusion cross sections are  obtained  by averaging the contributions
 from the different  orientation angles  $\alpha_1$ and $\alpha_2$ (relatively to the beam direction)
 of the  projectile and target nuclei, respectively \cite{NasirovNPA759}:
 \begin{eqnarray}
\label{averfus}
<\sigma_{\rm fus}(E_{c.m.},\ell)>&=&
\int_0^{\pi/2}\sin\alpha_1\int_0^{\pi/2}\sigma_{\rm fus}
(E_{\rm c.m.},\ell; \alpha_1,\alpha_2) \nonumber\\
&\cdot& \sin\alpha_2 d\alpha_1d\alpha_2.
 \end{eqnarray}

The partial fusion cross section $\sigma_{\rm fus}(E_{\rm c.m.},\ell; \alpha_1,\alpha_2)$ is
determined by the product of the partial capture cross
section $\sigma_{\rm cap}(E_{\rm c.m.},\ell; \alpha_1,\alpha_2)$
and   probability $P_{\rm CN}$ of the transformation of DNS
into CN:
 \begin{equation}
 \label{parfus}
 \sigma_{fus}(E,\ell;\alpha_1,\alpha_2)= \sigma_{cap}(E,\ell;\alpha_1,\alpha_2)
 P_{\rm CN}(E,\ell;\alpha_1,\alpha_2).
 \end{equation}

The capture probability and the largest value of orbital angular momentum ($\ell_d$),
 leading to capture at the given values of the orientation angles $\alpha_1$ and
$\alpha_2$, are  calculated  by  solving  equations of the relative motion
  of nuclei.
The calculation of capture and fusion cross sections were performed in the framework
 of the DNS model.  The details of this model can be found in Refs.
 \cite{FazioMPL2005,NasirovNPA759,FazioEPJ2004,NasirovPRC79}.

\subsection{Including surface vibration of spherical nucleus}

The projectiles $^{48}$Ca and $^{64}$Ni are nearly spherical nuclei.
In calculations of capture and fusion cross sections
the fluctuation of their shape around the
spherical shape. The nuclear radius,
 $R_P(\theta)=R_P^{(0)}(1+\sum_{\lambda=2,3}\beta^P_{\lambda}Y_{\lambda 0}(\theta))$,
 is parameterized in terms of vibrational amplitudes $\beta^P_{\lambda}$
to take into account the zero-point motion resulting from  the quadrupole and
octupole excitations. The surface vibrations are assumed to be independent
harmonic vibrations and  the distribution of the nuclear radius is
considered as a gaussian \cite{Esbensen81},
\begin{equation}
g(\beta^{(P)}_2,\beta^{(P)}_3)=\exp\left[-\frac{(\sum_{\lambda}\beta^{(P)}_{\lambda}Y^*_{\lambda0}
(\alpha_1))^2}{2{\sigma_{\beta_P}}^2}\right]\left(2\pi \sigma_{\beta_P}^2\right)^{-1/2} ,
\end{equation}
 where $\alpha_1$ is the direction of the symmetry axis
of the projectile shape when it has prolate ($\beta^{(P)}_2>0$)
or oblate ($\beta^{(P)}_2<0$) deformation.
It is assumed  $\alpha_1=0$ in our calculations.
The standard deviation of the gaussian distribution is calculated by the formula \cite{BohrMott}
\begin{equation}
\sigma_{\beta_P}^2=R_0^2\sum_{\lambda}\frac{2\lambda+1}{4\pi}
\frac{\hbar}{2D_{\lambda}\omega_{\lambda}}=\frac{R_0^2}{4\pi}\sum_{\lambda}\beta^2_{\lambda},
\end{equation}
 where $R_0\equiv R_P^{(0)}$; $\omega_{\lambda}$ is the frequency and $D_{\lambda}$ is the mass
 parameter of a collective mode. The full amplitude of the zero point motion is determined
 by formula \cite{BohrMott}
\begin{equation}
\beta_{\lambda}^2=\frac{(2\lambda+1)\hbar}{2D_{\lambda}\omega_{\lambda}}.
\end{equation}
We use $\lambda=2,3$ for  $^{48}$Ca and $^{64}$Ni projectiles.
The deformation parameters $\beta_{\lambda}$ related to the
 experimental values of the reduced electric quadrupole transition
 probability, $B(E\lambda)$,  of the first vibrational states $2^+$ and $3^-$
are taken  from Refs.\cite{Raman,Spear} (see Table \ref{Table1}), respectively.

\begin{table}[ht]
\label{Table1}
\centering
\caption{\small The quadrupole and octupole deformation parameters used for the first
excitation states for   $^{48}$Ca and  $^{64}$Ni nuclei.}
\begin{tabular}{|c|c|c|}
  \hline
  Nucleus & $\beta_2$\cite{Raman} & $\beta_3$ \cite{Spear}\\
  \hline
  $^{48}$Ca& 0.101  & 0.250 \\
  $^{64}$Ni & 0.179 &  0.230\\
  \hline
\end{tabular}
\\
\end{table}

We then average  the capture and fusion cross sections
over the values of the shape parameters used in the calculations:
\begin{eqnarray}
\label{vibaver}
<\sigma_{i}(E_{\rm c.m.},\alpha_2)>&=&
\int_{-\beta^{(0)}_{2}}^{\beta^{(0)}_{2}}
\int_{-\beta^{(0)}_{3}}^{\beta^{(0)}_{3}}\sigma_{i}
(E_{\rm c.m.}; \beta^{(P)}_2,\beta^{(P)}_3,\alpha_2) \nonumber\\
&\cdot& g(\beta^{(P)}_2,\beta^{(P)}_3) d\beta^{(P)}_2 d\beta^{(P)}_3,
\end{eqnarray}
with $i={\rm cap},\, {\rm fus}$ and with the weight function \cite{Esbensen81}

The results obtained by (\ref{vibaver}) were used in the following formula
 \begin{equation}
 \label{totalER2}
<\sigma_{\small ER}(E_{\rm c.m.})>=
\int_0^{\pi/2}\sigma_{\small ER}
(E_{\rm c.m.}; \alpha_2)\sin\alpha_2 d\alpha_2
 \end{equation}
to calculate the ER cross section by averaging only over  the different orientation
 angles of the symmetry axis $\alpha_2$ of the deformed target--nucleus.
The fusion excitation function is  determined by product of the partial
capture cross section $ \sigma_{\rm cap}(E_{\rm c.m.},\ell)$ and the fusion probability
 $P_{CN}$ \cite{Nas2007,GiardinaConf} of DNS at the various $E_{\rm c.m.}$ values:
 \begin{eqnarray}
 \label{totfus}
 \sigma_{\rm fus}(E_{\rm c.m.};\beta_P, \alpha_2)&=&\sum_{\ell=0}^{\ell_f}(2\ell+1)
 \sigma_{\rm cap}(E_{\rm c.m.},\ell;\beta_P, \alpha_2)\nonumber\\
 &\cdot& P_{\rm CN}(E_{\rm c.m.},\ell; \beta_P, \alpha_2).
 \end{eqnarray}

 For simplicity hereafter, we use $\beta_P=\{\beta^{(P)}_2,\beta^{(P)}_3\}$
  to  characterize the parameters of the first collective vibrational states
 2$^+$ and  3$^-$, respectively, in formulas of cross sections.

\subsection{Quasifission and fast fission of mononucleus evolving to equilibrium shape of compound nucleus}
\label{ffission}

Another binary process which leads to the formation of two fragments
similar to that from fusion-fission or quasifission is the fast fission.
According to the liquid drop model, the fast fission occurs only at large values of  angular momentum
$\ell > \ell_f$ causing disappearance of the macroscopic fission barrier $B_f(\ell)$ of the
 rotating nucleus  \cite{Sierk}.
 It is the  disintegration of the fast rotating mononucleus,  which
 survives  against quasifission (the decay of the DNS into two fragments without formation of the CN),
into two fragments.

In the case of very heavy nucleus  ($Z >$ 106), the fission barrier providing
their stability against fission appears only due to shell effects in their binding energy
\cite{ArmbrusAnnRevNP,Sobiczewski2007}. The damping of the shell effects decreases
the possibility of mononucleus to reach the CN equilibrium shape, and the mononucleus breaks down into
two fragments without reaching of CN shape. The fission barrier consists  of the
contributions of the macroscopic and microscopic parts.
The dependence of the fission barrier $B_{\rm fis}$ including shell correction $\delta  W$
on the critical angular momentum $\ell_f$ can be determined  by the formula
 \begin{equation}
 \label{fissb} B_{\rm fis}(\ell,T)=c \ B_{\rm fis}^{m}(\ell)-h(T) \ q(\ell) \ \delta W.
 \end{equation}
There is no macroscopic barrier ($B_{fis}^{m}$=0) for CN formed in these reactions under discussion.
The microscopic (shell) correction to the fission barrier $\delta W = \delta W_{sad} - \delta W_{gs} \simeq - \delta W_{gs}$
 is  taken from the table \cite{MolNix,Muntian03,KowalPRC,KowalIJMP}.
 The damping of the microscopic fission barrier on the excitation energy
and angular momentum of CN is taken into account  by formulae used in Ref. \cite{MandaglioPRC86}

 \begin{equation}
 \label{dampT}
 h(T) = \{ 1 + \exp [(T-T_{0})/d\}^{-1},
 \end{equation}
and

 \begin{equation}
 \label{dampL}
 q(\ell) = \{ 1 + \exp [(\ell-\ell_{1/2})/\Delta \ell]\}^{-1},
 \end{equation}
 where in  the formula (\ref{dampT}) $d=0.3$ MeV and $T_0$ = 1.16 MeV, and
 in  the formula (\ref{dampL}) $\ell_{1/2} = 20 $ and $\Delta \ell = 3 $
 values of parameters were used.

The part of the partial fusion cross section
 with $\ell>\ell_f$ is considered as partial fast fission cross section.
We should stress that for the superheavy elements
$\ell_f$ is not relevant quantity because there is no barrier connected with
the liquid drop model. Therefore, in this work we use  $\ell_f$ connected
 only with the shell corrections.

The fast fission cross section  is calculated
by summing the  contributions of the partial waves corresponding to the
range $\ell_f\le\ell\le\ell_d$ leading to the formation of the mononucleus:
 \begin{eqnarray}
 \label{fasfiss}
 \sigma_{\rm ff}(E_{\rm c.m.};\beta_P,\alpha_2)&=&\sum_{\ell_f}^{\ell_d}(2\ell+1)
 \sigma_{\rm cap}(E_{\rm c.m.},\ell; \beta_P, \alpha_2)\nonumber\\
 &\cdot& P_{\rm CN}(E_{\rm c.m.},\ell; \beta_P, \alpha_2).
 \end{eqnarray}

The capture cross section in the framework of the DNS model
is equal to the sum of the quasifission,
fusion-fission, and fast fission cross sections:
 \begin{eqnarray}
 \label{capture}
 \sigma^{\ell}_{\rm cap}(E_{\rm c.m.};\beta_1, \alpha_2)&=&
 \sigma^{\ell}_{\rm qfiss}(E_{\rm c.m.};\beta_1, \alpha_2) \nonumber\\
& +&\sigma^{\ell}_{\rm fus}(E_{\rm c.m.}; \beta_1, \alpha_2)\nonumber\\
 &+& \sigma^{\ell}_{\rm ffis}(E_{\rm c.m.}; \beta_1, \alpha_2).
 \end{eqnarray}
It is clear that the fusion cross section includes the cross sections of ERs
and fusion-fission products.

 Obviously, the quasifission cross section is defined by
\begin{eqnarray}
 \label{totqfis}
 \sigma_{qfis}(E_{\rm c.m.}; \beta_P, \alpha_2)&=&
 \sum_{\ell=\ell_f}^{\ell_d}(2\ell+1)\sigma_{cap}(E_{\rm c.m.},\ell; \beta_P, \alpha_2) \nonumber\\
 &\cdot& (1-P_{CN}(E_{\rm c.m.},\ell; \beta_P, \alpha_2)),
 \end{eqnarray}
 {\it i.e} the quasifission process can take place in the whole range of
 the orbital angular momentum values leading to capture, including central
collisions ($\ell=0$). This is important conclusion
 since the separation of the ranges of the angular momentum corresponding to
 the fusion-fission and
quasifission products by some critical value $\ell_{\rm cr}$
in the analysis of the angular distribution of the fissionlike products is doubtful.
The results of our calculations show that the ranges of the quasifission and
fusion-fission overlap.

\section{Comparison of the $^{48}{\rm Ca}$+$^{248}{\rm Cm}$ and
 $^{58}{\rm Fe}$+$^{232}{\rm Th}$ reactions leading to superheavy element  ${\rm Lv}$}
\label{CompLv}

\begin{figure}
\hspace{-0.5cm}\resizebox{0.5350\textwidth}{!}{\includegraphics{{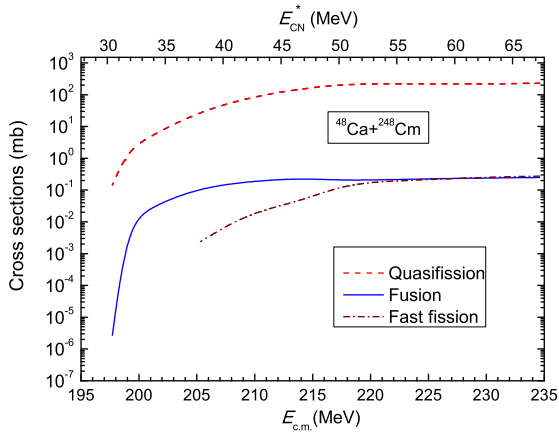}}}
\vspace*{-0.40 cm} \caption{\label{FusCaCm} (Color online) Capture (thin dashed line),
quasifission (thick dashed line),  fast fission (dot-dashed line) and fusion (solid line)
cross sections calculated by the DNS model for the $^{48}$Ca+$^{248}$Cm reaction.
 The excitation energy $E^*_{\rm CN}$ of CN (top axis)
is calculated by the use of the  M\"oller and Nix mass table \cite{MolNix}.}
\end{figure}

\begin{figure}
\vspace*{-0.15cm}
\hspace{-0.5cm}\resizebox{0.53\textwidth}{!}{\includegraphics{{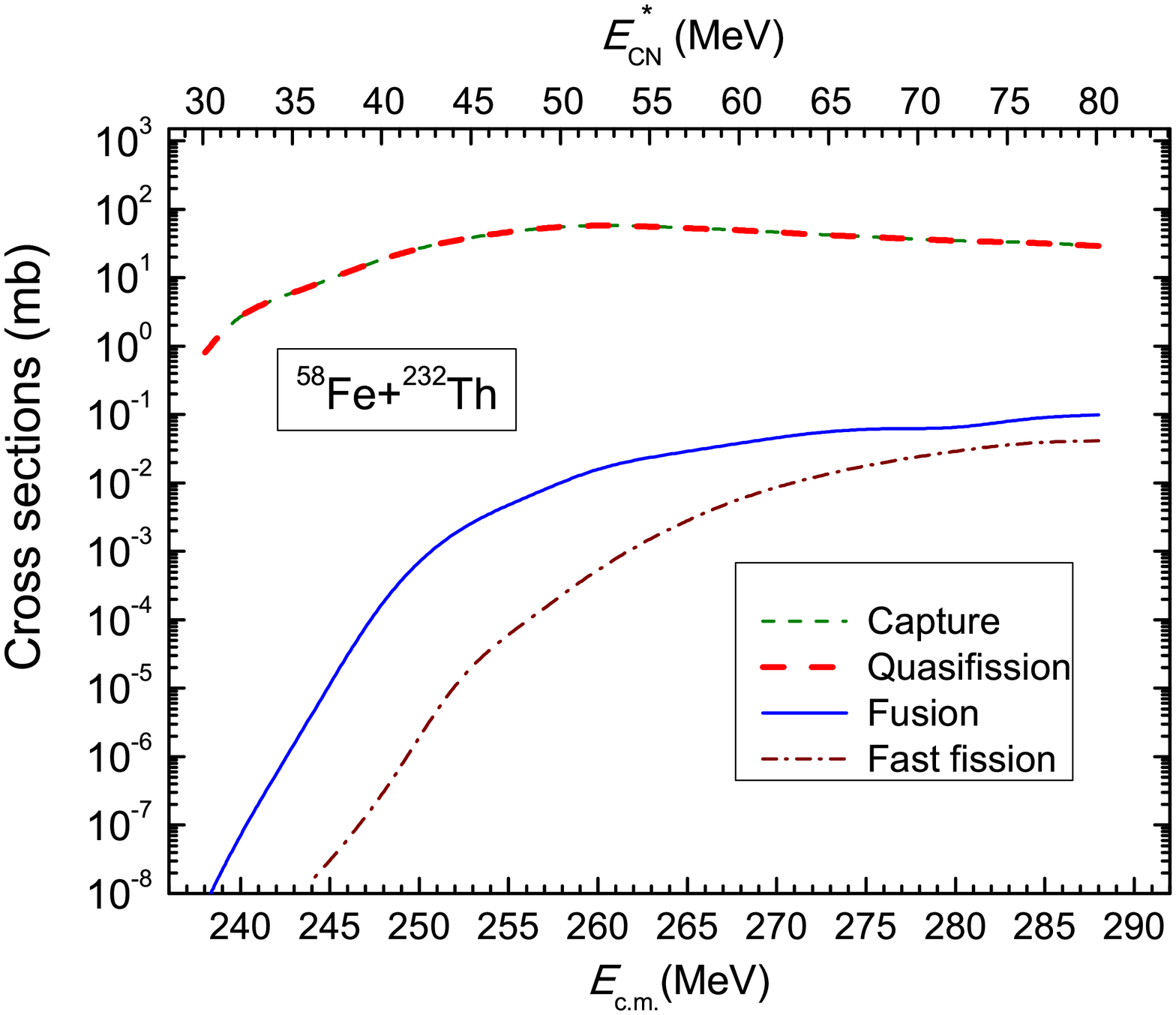}}}
\vspace*{-0.50 cm}
\caption{\label{FusFeTh} (Color online) Same as Fig. \ref{FusCaCm}
but for the  $^{58}$Fe+$^{232}$Th reaction.
 }
\end{figure}

 The importance of the charge asymmetry in the CN formation is seen from the comparison
of the $^{48}$Ca+$^{248}$Cm and
 $^{58}$Fe+$^{232}$Th reactions leading to the $^{296}$Lv and $^{290}$Lv CN, respectively.
Fusion excitation functions, which have been obtained for these reactions, are
presented by solid line in Figs. \ref{FusCaCm} and \ref{FusFeTh}, respectively.
We stress two main differences: 1) the cross section of the CN formation
is larger in the reaction with $^{48}$Ca in comparison with the $^{58}{\rm Fe}$+$^{232}{\rm Th}$
reaction; 2)  CN formed in the former reaction has smaller excitation energy.

  In Figs. \ref{FusCaCm} and \ref{FusFeTh}, capture and quasifission cross sections are nearly equal
    to each other  due to the dominant contribution
  of the quasifission cross section to capture in comparison with the sum of the fast fission and
  complete fusion.

 The number of events going to
quasifission increases drastically with the increasing
Coulomb interaction and rotational energy in the entrance channel
\cite{GiaEur2000,FazioPRC2005}. The Coulomb interaction increases
 with the increasing  charge number of the projectile or target nucleus,
and also it increases  with the decreasing  charge asymmetry of
colliding nuclei at fixed total charge number of DNS.
  The advantage of the $^{48}{\rm Ca}$+$^{248}{\rm Cm}$  in synthesis
of superheavy element is seen already in the CN formation stage.

\begin{figure}
\hspace{-0.5cm}\resizebox{0.57\textwidth}{!}{\includegraphics{{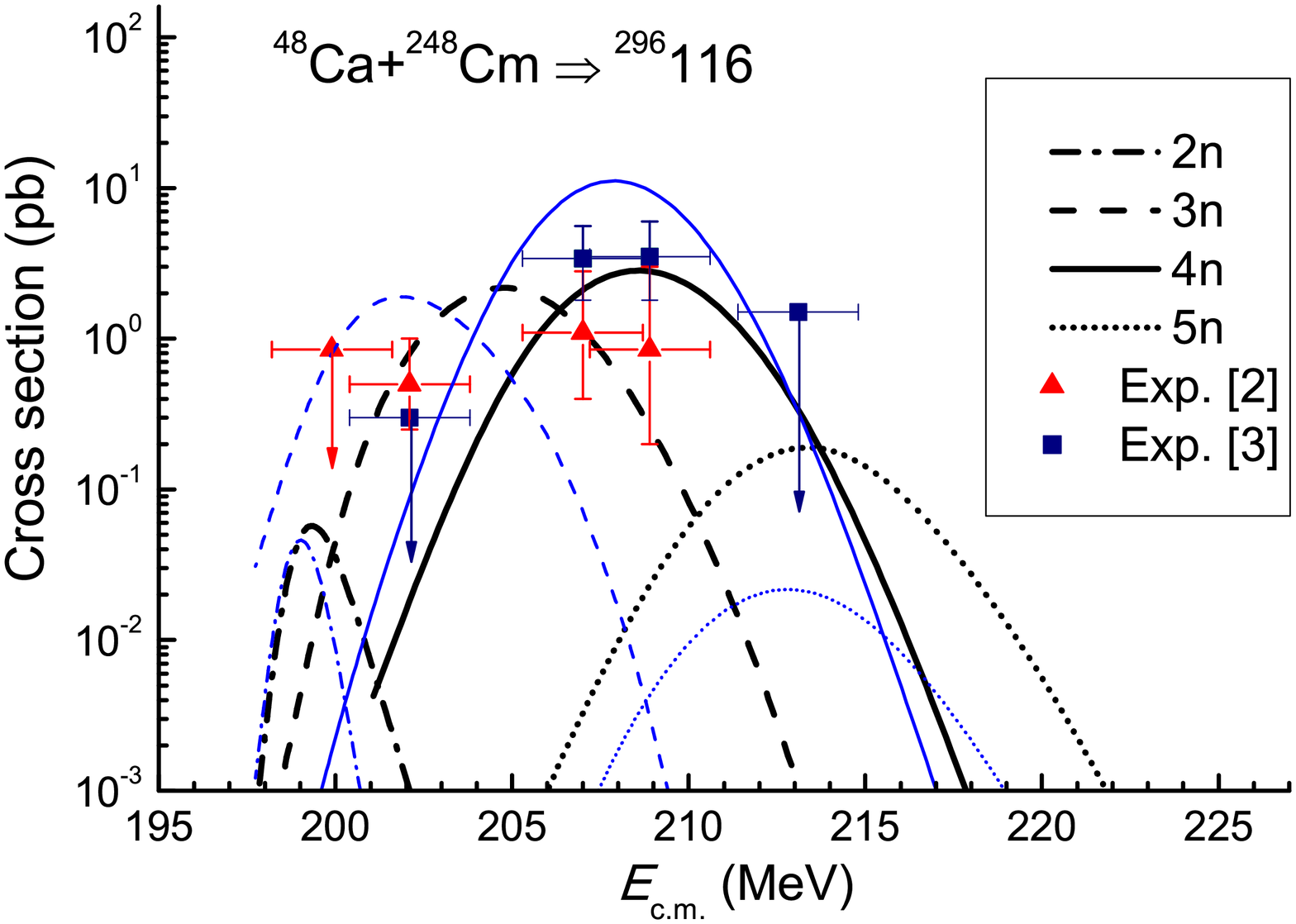}}}
\vspace*{-0.25 cm} \caption{\label{ERCaCm} (Color online)
Comparison between the ER excitation functions
for the $^{48}$Ca+$^{248}$Cm reaction calculated by using mass
tables of  M\"oller and Nix \cite{MolNix} (thin
lines) and  of the Warsaw group \cite{Muntian03}
(thick lines) for the 2n (dot-dashed lines), 3n (dashed  lines), 4n
(solid lines), and 5n (dotted lines) channels calculated by
the advanced statistical model
\cite{ArrigoPRC1992,SagJPG1998}. The experimental
data  of Ref. \cite{Oganessian04,FLNR} for the 3n and 4n channels are presented
 by triangles and squares, respectively.}
\end{figure}
 \begin{figure}
\hspace{-0.5 cm}\resizebox{0.52\textwidth}{!}{\includegraphics{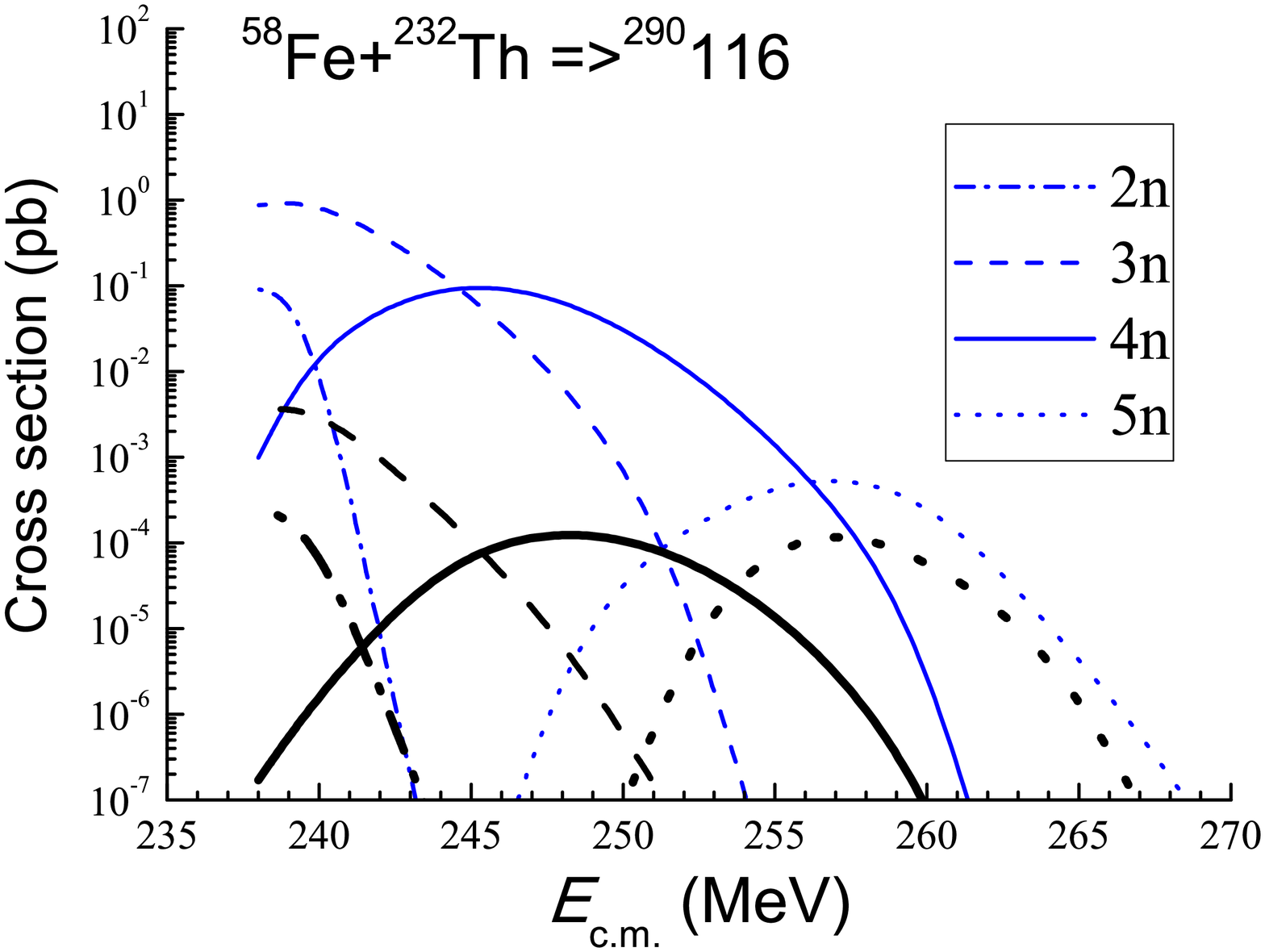}}
\vspace*{-0.30 cm} \caption{\label{ER58FeTh} (Color online)
The same as in Fig. \ref{ERCaCm} but for the $^{58}$Fe+$^{232}$Th reaction.}
\end{figure}

 One can see in Fig. \ref{FusFeTh} that the fusion excitation function
 decreases strongly at low energies  $E_{\rm c.m.}<$250 MeV. This
 effect is connected with the increase of hindrance to fusion since
  at these low energies the collisions
 with small orientation angles ($\alpha_1$-projectile and $\alpha_2$-target)
  can  only contribute to the CN formation \cite{NasirovNPA759,HindePCR74}.
  Collisions with larger orientation angles $\alpha_1$ and $\alpha_2$
  can not lead to capture since the collision energy in the center-of-mass system is not enough to
  overcome the Coulomb barrier of the entrance channel.
  For the DNS formed in collisions  with  small  orientation angles $\alpha_1$ and $\alpha_2$,
   the intrinsic fusion barrier $B^*_{\rm fus}$ is large  \cite{NasirovNPA759}.
So, the hindrance to complete fusion depends on the orientation angles:
 more elongated shape of the DNS formed  at collisions with small
 orientation angles (tip-to-tip  configurations)  promotes  the  quasifission  rather than
 the formation of the CN \cite{NasirovNPA759,HindePCR74}.

 Theoretical results of the ER cross  sections for the synthesis of the element $Z=116$ are
 compared  with the experimental data in Fig. \ref{ERCaCm}.  In this figure, the
 full triangles and squares show experimental data of the ER cross sections measured
 in 3n- and 4n-channels, respectively, in  the $^{48}$Ca+$^{248}$Cm reaction \cite{FLNR};
 the curves show theoretical results obtained in this work for the 2n-(dot-dashed line),
 3n-(dashed line), 4n-(solid line), and 5n-channel (dotted line)
 by the DNS and advanced statistical models \cite{ArrigoPRC1992,SagJPG1998} using  the mass tables
 of M\"oller and Nix \cite{MolNix} (thin lines) and  of Muntian {\it et al.}
\cite{Muntian03} (thick lines).

The mass values of the Warsaw group \cite{Muntian03} are larger than ones of
M\"oller and Nix \cite{MolNix}
  by  2-3 MeV for the isotopes of superheavy
 nuclei  with $Z > 110$ and fission barriers \cite{KowalPRC,KowalIJMP} are smaller by 2-3 MeV
in comparison with the similar values  of  M\"oller and Nix \cite{MolNix}.
 As a result,  the Warsaw group results lead to two main consequences:
 1) the excitation energy of the CN ($E^*_{\rm CN}=E_{\rm c.m.}+Q_{\rm gg}-V_{\rm rot}$)
will be lower since  the absolute  value of $Q_{\rm gg}=B_{\rm proj}+B_{\rm targ}-B_{\rm CN}$ (negative) is larger;
2) the fission probability will be large
 in comparison with the case of using fission barrier of the  M\"oller and Nix \cite{MolNix} model.
When the binding energies  and fission barriers of the  Warsaw group \cite{Muntian03} are used,
 the total score is that the survival probability $W_{\rm sur}$ becomes smaller
 in comparison with the case of using fission barrier of the
  M\"oller and Nix \cite{MolNix} model.

 The results of $\sigma_{\rm ER}$ for the $^{48}$Ca+$^{248}$Cm reaction,
 calculated by the use of the mass tables of the Warsaw group, better
describe the experimental data than the ones obtained by using M\"oller {\it et al}.
 The largest cross section of the yield of superheavy element
corresponds to the 4n-channel is about 10 pb  when the collision energy is in the range
 $E_{\rm c.m.}=$205--212 MeV  (thin solid line in Fig. \ref{ERCaCm}).

Concerning the ER formation in the more symmetric
$^{58}$Fe+$^{232}$Th reaction,  the results  indicate that
this reaction is less favorable in comparison with the $^{48}$Ca+$^{248}$Cm reaction
  to be used in the synthesis of superheavy element $Z=116$.
The largest cross section for the 3n-channel is about 1 pb (see Fig. \ref{ER58FeTh})
if we use the mass tables of M\"oller and Nix \cite{MolNix} (thin dashed lines),
while the use of the mass table and fission barriers of the Warsaw group leads
to the two orders of magnitude lower cross section (thick dashed line)
in comparison with the one in case of the using the mass tables of M\"oller and Nix.

\section{Comparison of $^{48}{\rm Ca}$+$^{249}{\rm Cf}$ and
 $^{64}{\rm Ni}$+$^{232}{\rm Th}$ reactions leading to superheavy element Z=118}
\label{CompCapFus}

\begin{figure}
\hspace{-0.5cm}\resizebox{0.535\textwidth}{!}{\includegraphics{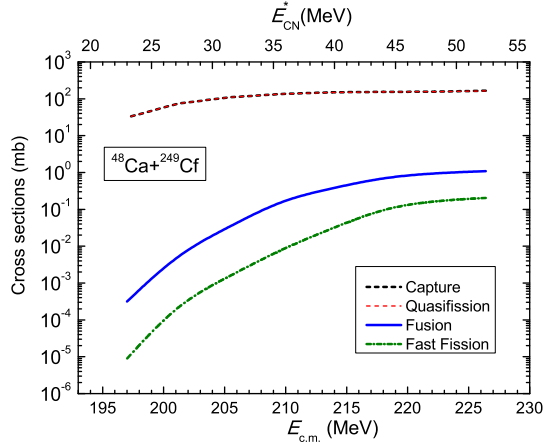}}
\vspace*{-0.20 cm} \caption{\label{FusCaCf} (Color online) Quasifission (dashed line),
fast fission (dot-dashed line), and complete fusion (solid line) excitation functions
calculated by the DNS model \cite{FazioMPL2005,NasirovNPA759,FazioPRC2005} for the
$^{48}$Ca+$^{249}$Cf reaction  which could lead to the $^{297}118$ CN.
 The capture cross section is not shown here because it is completely overlapped
  with the quasifission cross section.   The excitation energy $E^*_{\rm CN}$ (top axis)
is calculated by the use of the  M\"oller and Nix mass table \cite{MolNix}.}
\end{figure}

\begin{figure}
\hspace{-0.5cm}\resizebox{0.525\textwidth}{!}{\includegraphics{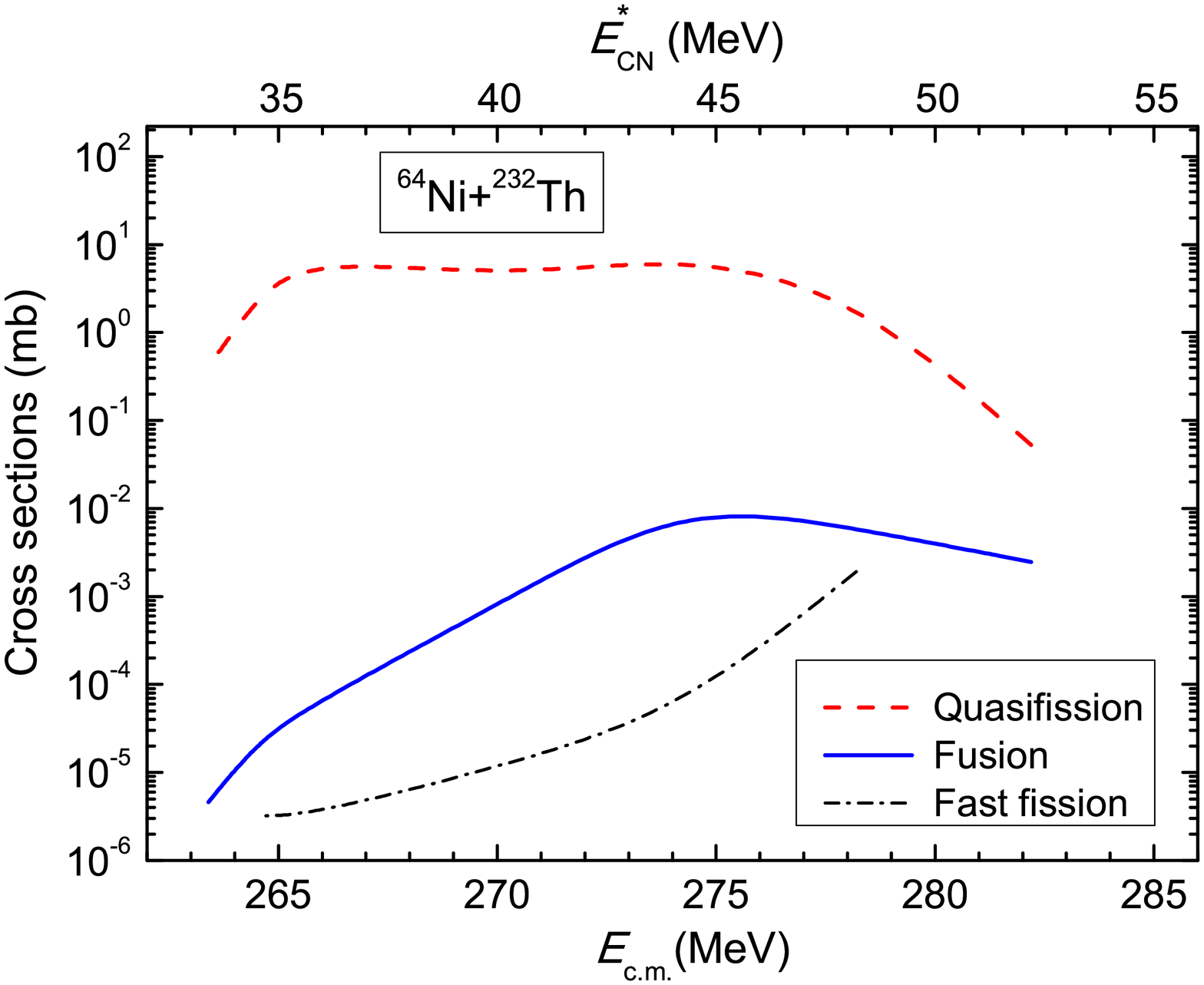}}
\vspace*{-0.35 cm} \caption{\label{FusNiTh} (Color online) Same as in Fig. \ref{FusCaCf}
but for the  $^{64}$Ni+$^{232}$Th reaction.}
\end{figure}

The dynamics of the capture and fusion for the
$^{48}{\rm Ca}$+$^{249}{\rm Cf}$ and $^{64}{\rm Ni}$+$^{232}{\rm Th}$ reactions
leading to  $Z=118$ differs significantly since the values of the Coulomb parameter
$z=Z_1\cdot Z_2/(A_1^{1/3}+A_2^{1/3})$ are very different,
197.47 and 248.41  for the reaction with Ca and Ni, respectively.
As shown in the systematic analysis \cite{GiardinaConf} at values $z>240$, the fusion
probability $P_{\rm CN}$ becomes less than 10$^{-7}$. The measured smallest value
of ER cross section is 50 fb \cite{RIKEN}
in the cold fusion  $^{70}$Zn+$^{209}$Bi reaction
 which has the Coulomb parameter $z=$247.62. It is very close to the value for
the reaction with Ni in our case.
The  capture, quasifission, fusion,  and fast fission cross sections calculated in this
work for the $^{48}{\rm Ca}$+$^{249}{\rm Cf}$ and
 $^{64}{\rm Ni}$+$^{232}{\rm Th}$ reactions are presented in Figs.
  \ref{FusCaCf} and  \ref{FusNiTh}, respectively.

\begin{figure}
\hspace{-0.50cm}\resizebox{0.525\textwidth}{!}{\includegraphics{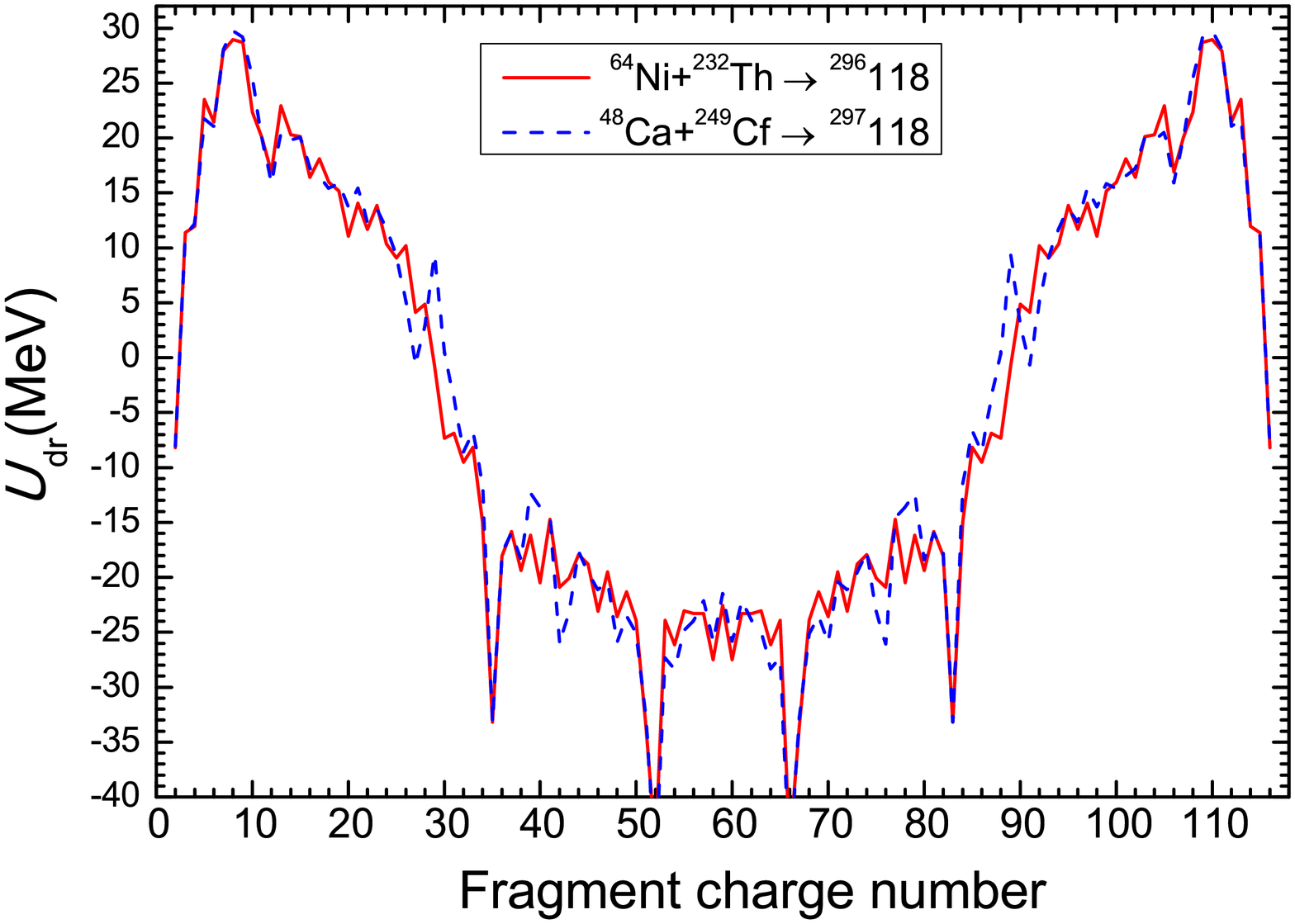}}
\vspace*{-0.50 cm} \caption{\label{DrivComp2} (Color online) Comparison of the driving
potentials calculated for the DNS formed in the $^{48}$Ca+$^{248}$Cf and $^{64}$Ni+$^{232}$Th
reactions which can lead to formation of isotopes $A$=297 and 296 of new superheavy element
$Z$=118 as a function of the fragment charge number.}
\end{figure}

\begin{figure}
\hspace{-0.5cm}
\resizebox{0.525\textwidth}{!}{\includegraphics{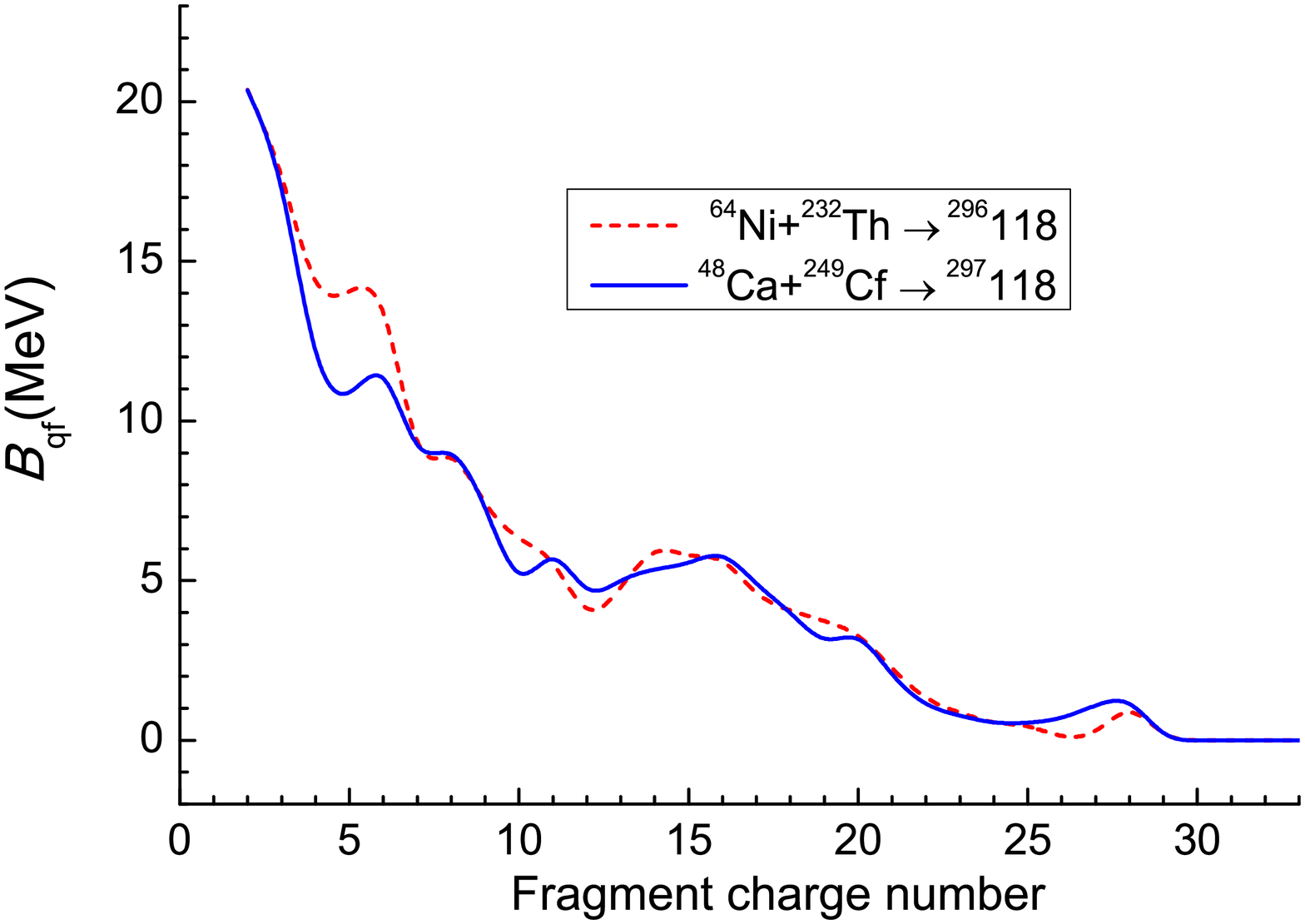}}
\vspace*{-0.60 cm} \caption{\label{BqfNiCa} The quasifission barriers of the
DNS fragments as a function of their charge numbers for the $^{48}{\rm Ca}$+$^{249}{\rm Cf}$
(solid curve) and  $^{64}{\rm Ni}$+$^{232}{\rm Th}$ (dashed curve) reactions.}
\end{figure}

  As we have stressed in Section \ref{CompLv}, in these reactions
    with massive nuclei, the capture and quasifission cross sections are equal
      since the sum of the fast fission and
  complete fusion is much smaller than the quasifission cross section.

  The comparison  between these figures shows that, at low energies,
 the capture cross section in the $^{64}$Ni+$^{232}$Th reaction
 is much smaller than  that in the $^{48}$Ca+$^{249}$Cf
 reaction. The curve of capture cross section for the   $^{64}$Ni+$^{232}$Th reaction
 goes down at larger energies $E_{\rm c.m.}>263$ MeV. But fusion cross section
 is very small due to drastic hindrance to complete fusion since at low
 energies only collisions with small values of orientation angles $\alpha_2$
 lead to the capture, {\it i.e} to formation of the long living DNS.
 DNS formed in collisions with small values of  $\alpha_2$  has large intrinsic barrier
 $B^*_{\rm fus}$ (see Fig. \ref{DrivComp2}) and small quasifission barrier $B_{\rm qf}$
 for this reaction in comparison with $^{48}$Ca+$^{249}$Cf (see Fig. \ref{BqfNiCa}).
This means that the potential well in the nucleus-nucleus interaction is
shallow  since the Coulomb interaction is stronger for more symmetric reactions
  ($z$=248.41 for $^{64}$Ni+$^{232}$Th reaction).
  These two reasons, which are  unfavorable for complete fusion \cite{NasirovNPA759},
   lead small  fusion probability for the $^{64}{\rm Ni}$+$^{232}{\rm Th}$ reaction
   at low energies.
  It is well known that, in collisions of nuclei at large orientation angles $\alpha_2$
  of symmetry axis of $^{232}$Th, the capture is possible  only at large
   energies $E_{\rm c.m.}>276$ MeV but due to the shallowness  of
   the potential well only small partial waves (small values of $\ell$)
   could lead to capture. The potential well disappears by the increase of
   (small values of $\ell$) at the large values of orientation angles  $\alpha_2$.
   This reason (the shallowness  of the potential well)
   leads to the decreasing the contribution from collision of nuclei into capture
   with the small values of $\alpha_2$ at large energies since
   the dissipation of the relative motion is not enough to cause
   trapping of the collision path into potential well. This case is
   demonstrated in Fig. \ref{TrapWell}b in Introduction.

 The advantage of the charge asymmetric system in complete
  fusion appears at the second stage (fusion) of the
 reaction mechanism leading to the ER formation.
 It is well known that the hindrance to complete fusion decreases with the increasing  DNS
 charge asymmetry. At the same time the DNS quasifission barrier, $B_{\rm qf}$,
 increases since the Coulomb repulsion forces decrease  with the decrease of the product
 $Z_1\cdot Z_2$.

 \begin{figure}
\vspace*{-0.5cm} \resizebox{0.525\textwidth}{!}{\includegraphics{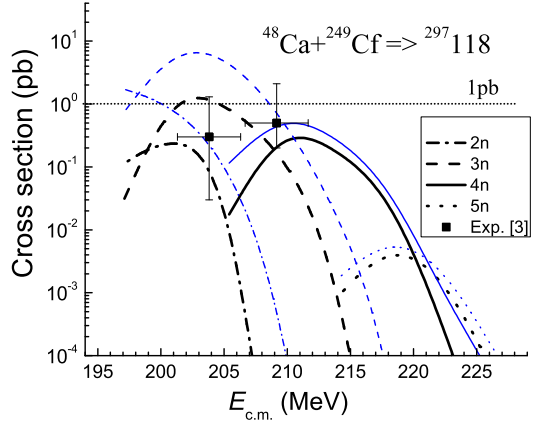}}
\vspace*{-0.4 cm} \caption{\label{ERCa249Cf} (Color online)
Comparison between the ER excitation functions
for the $^{48}$Ca+$^{249}$Cf reaction calculated by using mass
tables of  M\"oller and Nix \cite{MolNix} (thin
lines) and  of the Warsaw group \cite{Muntian03}
(thick lines) for the 2n (dot-dashed lines), 3n (dashed lines), 4n
(solid lines), and 5n (dotted lines) channels calculated by
the advanced statistical model
\cite{ArrigoPRC1992,SagJPG1998}. The experimental
data  of Ref. \cite{FLNR} are presented
 by squares.}
\end{figure}

 \begin{figure}
\hspace*{-0.5cm} \resizebox{0.525\textwidth}{!}{\includegraphics{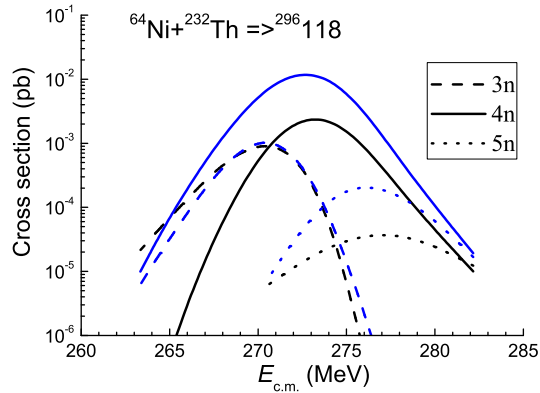}}
\vspace*{-0.40 cm} \caption{\label{ERNiTh} (Color online)
Same as in Fig. \ref{ERCa249Cf} but for the  $^{64}$Ni+$^{232}$Th reaction.}
\end{figure}

 The theoretical ER excitation functions  which can be formed in different
  neutron-emission channels for these two systems are presented in Figs.
 \ref{ERCa249Cf} and \ref{ERNiTh}. In each of the figures the evaporation  residue
 cross  sections for the neutron-emission channels obtained by using  binding energies and fission barriers
  calculated in the microscopic-macroscopic  models of  M\"oller and Nix \cite{MolNix} and
 of the Warsaw group \cite{Muntian03} are compared.

In Fig. \ref{ERCa249Cf},  we present the results for the $^{48}$Ca+$^{249}$Cf
reaction  \cite{MandaglioPRC86} leading to $^{297}$118 CN to compare with the results calculated
for the $^{64}$Ni+$^{232}$Th reaction leading to the same superheavy
element $^{296}$118 with smaller mass number.
In the $^{48}$Ca+$^{249}$Cf experiment (see Ref. \cite{FLNR})
 the superheavy element $^{294}$118 was observed after emission
of three neutrons  from the $^{297}$118 CN, at two projectile energies
corresponding to $E^{*}_{\rm CN}$=29.2 and 34.4 MeV.
 As seen in Fig. \ref{ERCa249Cf}
 the maximum values of cross sections connected with the 2n, 3n, and 4n
emission channels are in the range between 0.3 pb and 1.2 pb.

One can see from Fig. \ref{ERNiTh} that
the results obtained for all $x$n-channels of ER formation
  in the $^{64}$Ni+$^{232}$Th reaction are very small.
 The largest point of the 4n-channel curve is about 10 fb
 in the case when we used  masses of the M\"oller and Nix table \cite{MolNix}
 in our calculations. The use of the
That means this reaction is not favorable to observe
an event corresponding to the synthesis of any
isotopes of the superheavy element 118 by the modern
detectors.

The ER cross sections of the 4n and 5n channels for the $^{64}$Ni+$^{232}$Th reaction
obtained by the using masses of the M\"oller and Nix table \cite{MolNix} are larger
than those calculated by the using mass tables of Warsaw group \cite{Muntian03},
while the same behaviour does not appear by an evident way for the 3n channel.
 This is due to the effect of the different $E^*_{\rm CN}$ threshold
 values for the 3n channel when the masses of table \cite{MolNix} and Warsaw group \cite{Muntian03}
 are considered. In the first case the threshold energy for the 3n channel is
 at the $E^*_{\rm CN}=$35 MeV, while for the second case it is at $E^*_{\rm CN}\equiv$30.5 MeV.
 Due to the above-mentioned conditions for the  $^{64}$Ni+$^{232}$Th reaction,
  the excitation function of the 3n channel is strongly different in comparison
  with the ones shown in Fig. \ref{ERNiTh} for the 4n and 5n channels.

\section{Conclusions}

In the framework of the combined DNS and advanced statistical
models, the ER excitation functions have been calculated for the
$^{48}$Ca+$^{248}$Cm and $^{58}$Fe+$^{232}$Th reactions
leading to new superheavy element Lv ($Z$=116) and the
$^{48}$Ca+$^{249}$Cf and $^{64}$Ni+$^{232}$Th reactions
leading to new superheavy element $Z$=118.
The results obtained in this work are compared with
the experimental data given in  Refs. \cite{Oganessian04,FLNR}.
The ER excitation functions  of the 3n- and 4n-channels
for the $^{48}$Ca+$^{248}$Cm reaction  are  well described when
the Warsaw  group  mass
tables \cite{Muntian03} are used,  while  in both cases the use  of the  M\"oller and Nix
\cite{MolNix} mass tables leads to overestimation of the experimental data.

The theoretical values of ER cross sections for the $^{48}$Ca+$^{249}$Cf
reactions (Fig. \ref{ERCa249Cf}) show that the maximum values
of cross sections connected with the 2n, 3n, and 4n emission channels are in
the range between 0.3 pb and 1.2 pb when we use  the mass table and fission
barriers of Warsaw group. The experimental results about synthesis of
$Z$=118, which were presented in Ref. \cite{Oganessian04,FLNR}, are
in this range.

The results of the ER excitation functions for the $^{58}$Fe+ $^{232}$Th reaction
indicate that this reaction is less favorable to be used in the
synthesis of superheavy element $Z=116$.
The largest cross section for the 3n-channel is about 1 pb (see Fig. \ref{ER58FeTh}),
if we use the mass tables of M\"oller and Nix \cite{MolNix} (thin dashed line),
while the use of the mass table and fission barriers of Warsaw group leads
to much lower cross sections (thick dashed line), no more than 40 fb.
The modern state of the experimental setup does not allow to observe these events.

The results obtained for all $x$n-channels of ER formation
  in the $^{64}$Ni+$^{232}$Th reaction are very small (see Fig. \ref{ERNiTh}).
 The largest point of the 4n-channel curve is about 10 fb (thin solid line)
  when masses from the M\"oller and Nix table \cite{MolNix} were used
 in our calculations. The use of the Warsaw  group  mass
tables \cite{Muntian03} leads to values of the ER cross sections,
which are one order of magnitude lower than former case.
That means this reaction is not favorable to observe
an event corresponding to the synthesis of any
isotopes of the superheavy element 118 by the modern
detectors.

The reactions with the $^{232}$Th target and   $^{58}$Fe
and $^{64}$Ni projectiles can be useful to study the
role of the entrance channel and nuclear structure in formation of the fusion-fission
and quasifission products.
From the comparison of Figs. \ref{FusCaCm} and \ref{FusFeTh} we can state
that the fusion excitation function for
the $^{58}$Fe+$^{232}$Th reaction is from 10$^{-5}$ to  $10^{-2}$ orders
lower than the one calculated for the  $^{48}$Ca+ $^{248}$Cm reaction
 depending on the beam energy. The fusion cross sections of these
 compared reactions come closer at large collision energies
 but excitation energy $E^*_{\rm CN}$ of CN is so large
 and the ER cross sections of $x$n channels become negligibly small (see Figs.
 \ref{ERCaCm} and \ref{ER58FeTh})

 The strong hindrance at low energies
is caused by the large intrinsic fusion barrier $B^*_{\rm fus}$
for the small orientation angles of the symmetry axis of the deformed
projectile $^{58}$Fe ($\beta_2=0.2$) and target $^{232}$Th ($\beta_2=0.26$)
nuclei.
Therefore, the curve of the fusion excitation function increases slightly
with the increasing beam energy.

\textbf{Acknowledgments}
A. K. Nasirov is grateful to the Rare Isotope Science Project of
 the Institute for Basic Science of the Republic of Korea for the support
 the collaboration between the Dubna and Daejeon groups,
and he thanks the Russian Foundation for Basic Research for the partial
financial support in the  performance of this work.
The work of Y. Kim and K. Kim was supported by the Rare Isotope Science
 Project funded by the Ministry of Science, ICT and Future Planning (MSIP) and National Research Foundation
(NRF) of Korea.

\end{document}